# The response of a macromolecule near a tiled substrate

Pramod Kumar Mishra
Email: pkmishrabhu@gmail.com
*Department of Physics, DSB Campus, Kumaun University, Nainital (Uttarakhand) India-263002*

*Abstract—* We analyze response of a macromolecule near to a substrate; the substrate is tiled in the sequential and specific manner so that repeat units of the macromolecule may have different response on its adsorption in different directions onto the substrate. The lattice model of random walk has been used to analyze the adsorption-desorption behavior of an infinitely long homo-polymer molecule on a sequentially tiled substrate in three dimensions. The lattice model for the Gaussian chain and directed self-avoiding chain has been solved analytically. It has been emphasized on the basis of analytical estimates that a suitable coating may modify affinity of the macromolecule on the living surfaces and on the non living substrates in the complex manner which may be suitable means to control growth and also a route to restrict the spread of the deadly microbes.

*Keywords—* Macromolecule, Affinity, tiled substrate, analytical results

## 1. Introduction

It is well understood that coating on any substrate leads protection of the substrate [1-2] from its degradation [3-4], and also by means of the coating cream/chemicals on the living surfaces it may be possible to get different response on affinity of the dust particles or the microorganism towards the living surfaces. The microorganism may be made of specific hydrophobic macromolecules [5-6] and thus such microorganism may be easily attached to any organic surface due to hydrophobic nature of the molecule, while it may not be possible to attach the hydrophobic microorganism to any hydrophilic [5-6] substrate. On the other hand, it is possible to destroy the hydrophobic microbes by dissolving it in the hydrophilic solvents. Since, the virus which is covered in a hydrophobic cell may be easily attached to organic substrate and therefore it may be carried away from one site to another. If this virus is dangerous for the human, such virus may be cleaned using suitable organic compounds on the skin of the human and also by coating the non-living surface; On the other hand the non-living surface may be sterilized using suitable organic compounds.

In order to mimic the structure of microbes schematically, we consider homo-polymer chain in three dimensions and an open coil like structure of the chain has been treated as a macromolecule for the sake of mathematical simplicity. Such coil shaped macromolecule is assumed near to the sequentially tiled surface where the monomers have different possibility of attachment or polymerization along different directions on the tiled surface. The elastic behavior of the macromolecule has been incorporated into the model using the Boltzmann weight corresponding to each bend in the chain.

The model for the adsorption of a flexible homo-polymer chain had been widely studied [7-9]; and also there are several reports available in the literature regarding adsorption of the semi-flexible homo-polymer chain [8-13]. However, adsorption of semi-flexible homo-polymer chain on sequentially tiled surface is not reported. Therefore, we have chosen model of an ideal polymer chain in addition to a self-avoiding polymer chain to analyze adsorption of the chain on a sequentially tiled surface/substrate.

The paper is organized as follows: In the section two, we describe the lattice model for two and three dimensional polymer chain in brief; and the partition function of the two dimensional chain is obtained in the thermodynamic limit in the section three (A) for the cases when chain is in the two dimensional bulk as the Gaussian chain and also as a self-avoiding chain. In the section three (A), we used ideal chain as a random walk model, partially directed self-avoiding and fully directed self-avoiding walk models. We have chosen partially and fully directed self avoiding walk models in three dimensions also to analyze the adsorption-desorption transition related results on tiled surface in the section three (B). We summarize our results and conclude the discussion in the section four.

## 2. The Model

We consider lattice model of an ideal polymer chain [7-9] where a site of the substrate may be visited more than once by the monomers of the chain because the monomers are assumed to have no volume in the Gaussian chain model [7-9]. Thus, in the case of two dimensional models of an ideal chain, $\pm x$ directions and $\pm y$ directions i. e. all directions are admissible by the walker in two dimensions while enumerating the conformations of an ideal chain on the tiled surface. One end of the chain is grafted onto the substrate (*i. e.* at a point O) for the all the lattice models that we have considered in the present manuscript.

We have also considered the case of the self-avoiding chain where partially directed and the fully directed walks are used to mimic conformations of the chain under partially directed self-avoiding walk (*PDSAW*) model and the fully directed self-avoiding walk (*FDSAW*) model, respectively [8-13]. In the case of square lattice (two dimensional case), the walker is allowed to take steps along $+x$ and $\pm y$ directions for the *PDSAW* model [8-9], (A walk of fifteen monomers are shown in the figure no. 1 for a two dimensional PDSAW); while only $+x$ and $+y$ directions are allowed to steps for the walker in the *FDSAW* model [8-9]. We have chosen three dimensional models for the self-avoiding polymer chain and a *PDSAW* and the *FDSAW* models have been solved analytically.

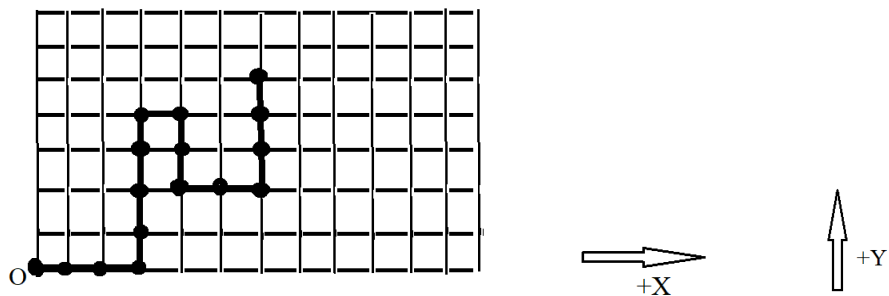

**Figure No. 1:** A partially directed self-avoiding walk of 15 monomers are schematically shown in this figure. There are six monomers along +x direction while there are seven monomers along +y direction and remaining two monomers are along –y direction. There are five bends shown in this walk and hence the Boltzmann weight of the fifteen monomers long PDSAW is $u^6 v^9 k^5$.

In the case of three dimensions (*i. e.* cubic lattice case), the walker is allowed to take steps along $+x$, $\pm y$ and $+z$ directions for the *PDSAW* model [8-9], while in the case of *FDSAW* model, walker's steps are allowed along $+x$, $+y$ and $+z$ directions [8-9]. The surface is two dimensional and it is sequentially and specifically tiled. Since, the substrate/surface is two dimensional for three dimensional model therefore, along $x$-direction a monomer may be added (polymerized) in the chain with effective probability $u$, while $v$ is the effective probability of polymerization/addition of a monomer along y-direction. The grand canonical partition of the chain may be written as,

$$G(u,v,k) = \sum_{P=1}^{N\to\infty} \sum_{K=0}^{N-P} \sum_{N_B=0}^{N-1} u^P v^K k^{N_B} \tag{1}$$

Here, $u$ and $v$ are the Boltzmann weight of the steps along $x$ (and $-x$) and $y$ (and $-y$) directions respectively; term $k$ is the Boltzmann weight corresponding to each bend in the chain conformations. The bending angle 90 degree corresponds to stiffness weight $k$ while in the case of ideal chain (*i. e.* the Gaussian chain) the Boltzmann weight is $k^2$ ($=K$) which corresponds to the bending angle 180 degree. While in the case self-avoiding walk model, either the bending angle is 90 degree or there is no bend in the chain segments.

### 3. The Calculations and the Results

We consider a homo-polymer chain on two dimensional substrate and we model the polymer chain on square lattice where the chain is ideal (*i. e.* the Gaussian chain) and all directions are permitted for the walker to take steps [8-9].

### A. Two Dimensional Bulk Case

In the case of two dimensional random walk models for an ideal chain on the square lattice the components of the partition function is written as,

$$X_{i(=1,2)}(u,v,k) = u + uX_i(u,v,k) + uKX_i(u,v,k) + 2ukY_j(u,v,k) \tag{2}$$

Here, $K=k^2$ and $X_i(u,v,k)$ is the sum of the Boltzmann weight of all walks having first step along $X_i$ direction ($X_1(u,v,k) = X_2(u,v,k)$); and the other components are written as,

$$Y_{j(=1,2)}(u,v,k) = v + vY_i(u,v,k) + vKY_j(u,v,k) + 2vkX_i(u,v,k) \tag{3}$$

Thus, in the case two dimensional model of an ideal chain ($Y_1(u,v,k) = Y_2(u,v,k)$); the partition function is written as,

$$Z^{2D}_{Ideal}(u,v,k) = \frac{2u(1-v-Kv) + 2v(1-u-Ku) + 8uvk}{(1-u-Ku)(1-v-Kv) - 4k^2 uv} \tag{4}$$

Thus, from the singularity of the partition function (i. e. Eqn. no. 4) the condition for the polymerization of an infinitely long ideal chain on the two dimensional substrate is written as follows:

$$u_c(v_c,k) = \frac{1 - v_c - Kv_c}{(1+K)(1-v_c-Kv_c) + 4k^2 v_c} \tag{5}$$

The variation of $u_c$ is shown in the Figure No. 2(i) for an ideal chain on two dimensional substrate case, where $u_c$ and $v_c \leq 0.5$ for $k=1$ and substituting $u_c=v_c$ for $k=1$, we have $u_c=1/4$ and it is a well known result of $u_c$ for the ideal chain on square lattice [8-9]; and similarly we may obtain the critical value of $u_c$ for two dimensional self-avoiding chain under the *PDSAW* and also for two dimensional *FDSAW* cases as follows:

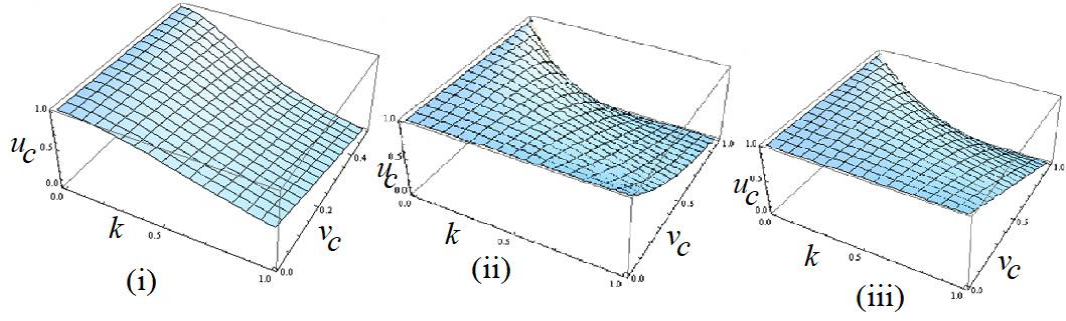

**Figure No. 2:** The polymerization of a macromolecule on tiled substrate is shown using the models of an ideal chain in the figure 2(i); and the polymerization of the real polymer chains are also shown in the 2(ii) for PDSAW model and in the figure 2(iii) for the FDSAW model. It has been shown that the possibility of the polymerization varies with stiffness of the chain, and it also varies with the probability of addition of another monomer along other possible direction for the addition/polymerization of monomer in the chain.

$$u_c(v_c,k) = \frac{1-v_c}{1+v_c(2k^2-1)} \tag{6}$$

$$u_c(v_c,k) = \frac{1-v_c}{1+v_c(k^2-1)} \tag{7}$$

The variation of $u_c$ as shown in eqn. no. (6) for *PDSAW* and by eqn. no. (7) for the *FDSAW*; and it is graphed in the figure no. 2(ii) and 2(iii) for *PDSAW* and *FDSAW*, respectively.

## B. Polymer adsorption on the tiled surface for three dimensional model

We consider the adsorption of an infinitely long self-avoiding polymer chain in three dimensions and the components of the partition function for partially directed self avoiding chain [8-13] using cubic lattice is written as,

$$X(u,v,k) = u + uX(u,v,k) + 2ukY(u,v,k) + ukZ(g,k) \tag{8}$$

$$Y(u,v,k) = v + vY(u,v,k) + vkX(u,v,k) + vkZ(g,k) \tag{9}$$

And

$$Z(g,k) = g + gZ(g,k) + ukY(g,k) + 2kgZ(g,k) \tag{10}$$

Here, $g$ is the fugacity of the monomer in the bulk (*i. e.* away from the surface) and effective fugacity $u$ and $v$ contains $g$ into itself; however, when the surface attraction is taken into consideration the surface components of the partition function may be written as,

$$S_x(w,u,v,k) = wu + wuS_x(w,u,v,k) + 2kwuS_y(w,u,v,k) + wukZ(g,k) \tag{11}$$

$$S_y(w,u,v,k) = wv + wvS_y(w,u,v,k) + kwvS_x(w,u,v,k) + wvkZ(g,k) \tag{12}$$

Where term $w$ is the Boltzmann weight corresponding to the monomer-surface attraction, we may solve the eqn. nos. (10-12) to obtain the partition function of a self-avoiding polymer chain in three dimensions where the monomers of the real chain have attractive interaction with the two dimensional specifically tiled surface. The adsorption transition point is the function of the monomer fugacity and the stiffness of the semi-flexible chain and the transition point for the partially directed self-avoiding polymer chain may be written as,

$$w_c(u_c, v_c, k) = \frac{(u_c + v_c) \pm \sqrt{(u_c + v_c)^2 - 4u_c v_c (1 + 2k^2)}}{2u_c v_c (1 + 2k^2)} \quad (13)$$

The variation of positive root of the transition point ($w_c$) with relevant parameters is shown in the figure no. 3(i-iii) for *PDSAW* model, provided $(u_c + v_c)^2 - 4u_c v_c (1 + 2k^2) \geq 0$. Similarly, one can obtain the adsorption transition point for the fully directed self-avoiding walk model and the transition point may be written as,

$$w_c(u_c, v_c, k) = \frac{(u_c + v_c) \pm \sqrt{(u_c + v_c)^2 - 4u_c v_c (1 - k^2)}}{2u_c v_c (1 - k^2)} \quad (14)$$

The transition points are shown in the figure no. 3(iv-vi) for the fully directed walk model of three dimensional case, provided $(u_c + v_c)^2 - 4u_c v_c (1 - k^2) \geq 0$; where adsorption of the chain is occurring on the specifically tiled surface.

We have plotted the results for the partially directed walk model in three dimensional space case in the figure no. 3(i), 3(ii) and 3(iii) which corresponds to the stiffness value *k=0.1, 0.5* and *0.9* respectively and in this model the flexible chain adsorption is not possible on the sequential substrate due to large entropy of the polymer chain. However, in the case of fully directed walk model we have chosen stiffness value of the chain *k=0.1, 0.5* and *0.99*.

### 4. The Summary And The Conclusions

The lattice model of an ideal chain [8-9] has been solved to estimate the possibility of polymerization of an infinitely long linear semi-flexible chain on a two dimensional specifically tiled surface and also lattice models of the partially and the fully directed self-avoiding walk models have also been solved to compare the results obtained for all three cases on the specifically tiled surface.

The lattice model of self-avoiding polymer chain of the partially and the fully directed walks [8-9] has also been solved on a cubic lattice to analyze the adsorption-desorption phase transition behavior of an infinitely long semi-flexible polymer chain on the tiled surface. The variation of the transition point has been shown for the admissible values of the monomer's effective fugacity and the stiffness of the chain. There are specific results available in the literature regarding adsorption of the molecule on surface [14-15].

Specifically tiled surface is assumed so that an edge of the walk which represents the monomer of the chain and the edge lying along *x*-direction have effective fugacity *u* and the monomer of the polymer which is along the *y*-direction has effect fugacity *v*. It is an artificial substrate which is decorated in such a manner such that it allows different fugacity along *x* and *y* directions. Because, such assumption allows us to solve model of the self-avoiding polymer chain analytically to investigate the adsorption-desorption transition behavior. It is evident from present investigation that the adsorption transition point's dependency on the stiffness of the chain is much evolved and complex rather than the dependence of the adsorption transition point on the stiffness of the chain for co-polymer chain case [11]. The idea of tiled substrate is originating from fact that the uniform substrate may be coated using the hydrophilic and the hydrophobic chemicals [5-6] on alternate arms of a square shaped plaque. Thus, the repeat units (monomers) of a macromolecule may have different affinity to bind with the tiled substrate. It will help to bind the hydrophobic and hydrophilic microbes and harmful virus may be destroyed on the substrate.

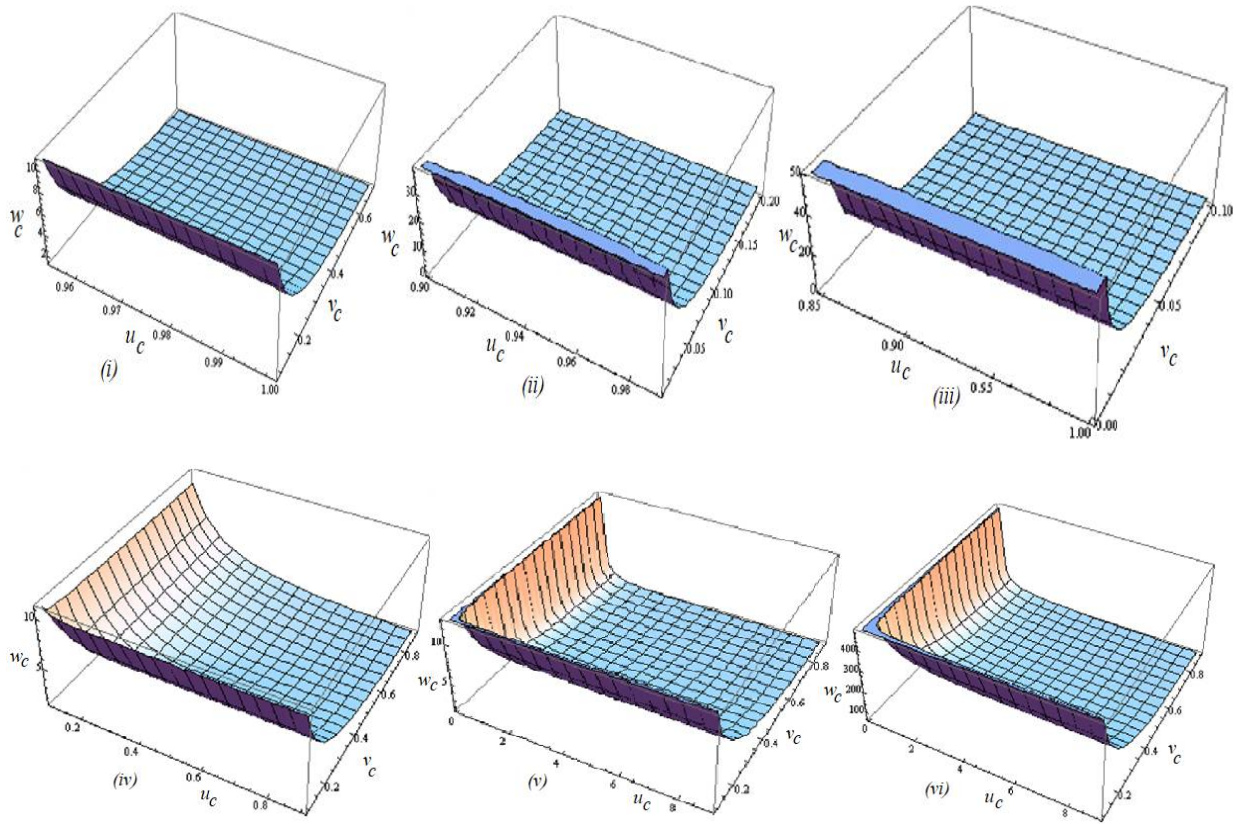

**Figure No. 3:** The adsorption transition point ($w_c$) is shown for three dimensional model of the self-avoiding polymer chain. The transition point for the partially directed self avoiding walk model is shown in the figure 3(i-iii) and also for the fully directed polymer chain in the figure 3(iv-vi) when the adsorption of the chain is taking place on the sequentially and specifically tiled two dimensional surface. The value of *k* for figure nos. 3(i)-3(vi) is 0.1, 0.5, 0.9, 0.1, 0.5 and 0.99 respectively.

It is expected that such approach may be an alternative to introduce defects in the polymer conformations and statistics of the defected polymer chain may also be analyzed for quench and annealed defect cases [16-17].

**About the Author: Dr. Pramod Kumar Mishra (**also know as **P. K. Mishra);**
Publons-id: http://publons.com/a/1336175/
Orcid id: http://orcid.org/0000-0003-4640-2374